\documentclass{article}
\usepackage{arxiv}
\usepackage[utf8]{inputenc} 
\usepackage[T1]{fontenc}    
\usepackage{hyperref}       
\usepackage{url}            
\usepackage{booktabs}       
\usepackage{amsfonts}       
\usepackage{nicefrac}       
\usepackage{microtype}      
\usepackage{lipsum}		
\usepackage{graphicx}
\usepackage{doi}
\usepackage{float}
\usepackage{array}
\usepackage{enumitem}
\usepackage{eurosym}
\usepackage[square,sort,comma,numbers]{natbib}

\title{The OAD Flagship Ecosystem}

\author{ \href{https://orcid.org/0000-0002-9745-0504}{\includegraphics[scale=0.06]{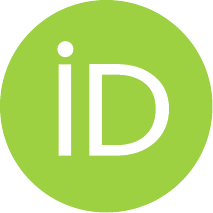}\hspace{1mm}Joyful E. Mdhluli\textsuperscript{1,2}}{ on behalf of the IAU Office of Astronomy for Development} \thanks{Visit our website, https://astrotourism.astro4dev.org/ or email: astrotourism@astro4dev.org} \\
	$^{1}$International Astronomical Union's Office of Astronomy for Development,\\
	$^{2}$South African Astronomical Observatory, Cape Town, 7925, South Africa\\
	\texttt{joy@astro4dev.org} \\
}

\renewcommand{\shorttitle}{\it{OAD Flagship Ecosystem}}

\hypersetup{
pdftitle={The OAD Flagship Ecosystem},
pdfsubject={astro-ph.IM},
pdfauthor={Joyful E. Mdhluli},
pdfkeywords={Astrotourism, Astronomy for Mental Health, Astronomy for Skills, Astronomy for Development},
}

\begin{document}
\maketitle

\begin{abstract}
The International Astronomical Union's Office of Astronomy for Development (IAU OAD) uses astronomy as a tool to address societal challenges and contribute to sustainable development. Building on more than a decade of project funding and implementation, the OAD has developed a portfolio of flagship projects that represent tested and scalable applications of Astronomy for Development across thematic areas including socio-economic development, science diplomacy, skills development, inequality reduction, and technology transfer. To support the growth and long-term sustainability of these initiatives, the OAD has established the Flagship Ecosystem, a framework built around four interconnected pillars: Resources, Training, Community, and Implementation.

This paper presents an overview of the OAD Flagship projects, the structure and components of the Flagship Ecosystem, and explores how it supports the translation of astronomy-based interventions into sustainable development outcomes. The ecosystem provides open-access resources, capacity-building opportunities, communities of practice, funding mechanisms, and evidence-generation activities that enable individuals and organizations to implement and scale astronomy-for-development initiatives. Grounded in principles of inclusivity, openness, sustainability, participation, and evidence-informed practice, the ecosystem aims to strengthen the global impact of Astronomy for Development while fostering collaboration across diverse sectors and regions.
\end{abstract}


\section{Introduction}       
Science for Development refers to the use of scientific research, technology, and innovation to address systemic challenges and improve quality of life, particularly in low- and middle-income countries. It leverages evidence-based approaches to promote economic growth, reduce poverty, strengthen food security, improve health outcomes, and address environmental and climate-related challenges. The field is underpinned by several key pillars, including evidence-based policy, capacity building, and the advancement of the United Nations Sustainable Development Goals (SDGs) (\cite{sdgs}). Through investments in research infrastructure, education, and local expertise, Science for Development seeks to empower communities to develop sustainable solutions to their own challenges.

Astronomy for Development (Astro4Dev) is an emerging field that sits within the broader Science for Development ecosystem. It seeks to use astronomy as a tool for addressing societal challenges through its infrastructure, technologies, knowledge, skills, and inspirational potential. While astronomy is often perceived as a highly specialized and seemingly distant scientific discipline, it offers unique opportunities to contribute to sustainable development. As a gateway science, astronomy can inspire interest in STEM, foster critical thinking, promote international collaboration, stimulate technological innovation, and strengthen cultural connections through our shared experience of the night sky. In this way, astronomy serves as more than a scientific endeavor - it acts as a bridge across disciplines, cultures, and technologies, making it a versatile tool for tackling global development challenges.

The International Astronomical Union's Office of Astronomy for Development (IAU OAD) was established to advance the use of astronomy as a tool for development by mobilizing the human and financial resources necessary to realize astronomy's scientific, technological, educational, and cultural benefits for society. This mission is primarily implemented through funding, coordinating, and supporting projects that use astronomy-based interventions to address sustainable development challenges. The OAD supports initiatives that contribute to one or more of the SDGs. Since its establishment, the OAD has funded and coordinated 254 projects implemented across 113 countries, with programs contributing to at least 13 of the 17 SDGs. These projects demonstrate the diverse ways in which astronomy can support education, skills development, economic growth, social inclusion, science diplomacy, and community well-being around the world.

\section{Thematic Areas}       
Part of the OAD’s strategy is to coordinate global flagship projects, which would be implemented on a much larger scale than those funded through the annual call for proposals, and funded through external fundraising. The OAD has identified five key themes that encapsulate the concept of astronomy for development. These themes were selected in early 2019 based on experience in supporting projects and reviewing more than 1,000 proposals. Input from the OAD regional offices was requested and special OAD projects, partnerships, and international trends were considered (\cite{themes}).

These themes represent some of the most effective and successfully tested applications of astronomy for development. The majority of projects funded through the annual call for proposals align with these themes. The five themes are:

\subsection{Theme 1: Sustainable, local socio-economic development through Astronomy (e.g. astrotourism, observatories for communities, etc)}
This theme aims to use an astronomical facility, such as an observatory or planetarium, as a “hub” to stimulate various socio-economic benefits for the local community. Benefits could include job creation through astronomy-related tourism; community skills development; educational programs; stimulation of local innovation; alternative activities for youth in order to keep away from negative/harmful activities; and infrastructure development. The establishment of such a facility in close collaboration with relevant government, industry, academic and development partners, local and traditional leadership, will ensure the sustainability of the initiative (\cite{socioeconomic}).

This aligns directly with several SDGs, including:
\begin{itemize}
    \item \textbf{SDG 4 (Quality Education):} Astronomy is a highly effective gateway science that inspires curiosity and engagement with STEM (Science, Technology, Engineering, and Mathematics).
    \item \textbf{SDG 8 (Decent Work and Economic Growth):} Astronomy creates employment opportunities, stimulates innovation, and supports emerging economic sectors.
    \item \textbf{SDG 11 (Sustainable Cities and Communities):} Astronomy contributes to preserving cultural heritage, protecting natural resources, and promoting sustainable tourism.
    \item \textbf{SDG 17 (Partnerships for the Goals):} Astronomy is inherently international and collaborative, making it an excellent platform for building partnerships especially between government, industry, academic and development partners, local and traditional leadership.
\end{itemize}

\begin{figure}[H]
    \centering
    \includegraphics[width=0.2\textwidth]{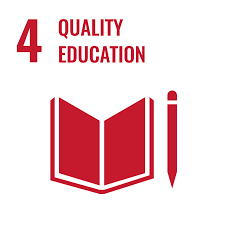}
    \includegraphics[width=0.2\textwidth]{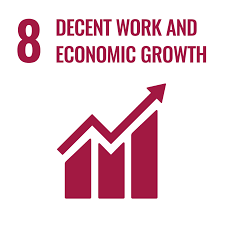}
    \includegraphics[width=0.2\textwidth]{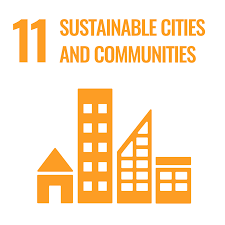}
    \includegraphics[width=0.2\textwidth]{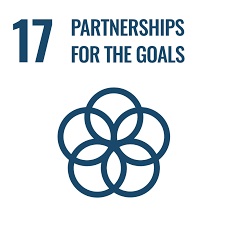}
    \caption{Sustainable Development Goals (SDGs) aligned with astrotourism. Image Credit: \href{https://sdgs.un.org/goals}{United Nations Sustainable Development Goals}\cite{sdgs}}
\end{figure}

\subsection{Theme 2: Science diplomacy through Astronomy: Celebrating our Common Humanity (e.g. peace, post-conflict, partnerships, policy, etc)}
Astronomy brings us a perspective of the beauty and scale of the universe. Most famously, astronomer Carl Sagan described the Earth as a “Pale Blue Dot”. He used the cosmic perspective to inspire and influence how people interact with their fellow human beings, and our planet. The goal of this theme is to use the inspirational themes of astronomy – namely, that we live under one sky and that we all have innate similarities on the basis of being human – to promote SDGs 4 (Quality Education), 13 (Climate Action), 10 (Reduced Inequalities) and 16 (Peace, Justice, and Strong Institutions).

\begin{itemize}
    \item \textbf{SDG4 (Quality Education):} Ensure inclusive and equitable quality education and promote lifelong learning opportunities for all. By inspiring people with the fascinating Universe, the projects under this flagship will attempt to increase the fraction of the population who are interested in following careers in science and technology and promote literacy in science and technology, particularly among the more vulnerable and disadvantaged communities. 
    \item \textbf{SDG10 (Reducing Inequalities within and among countries):} By encouraging a sense of global citizenship and unity we aim to shift people’s consumption patterns, also increasing support for sustainable policies that promote equality.  This flagship promotes the theme of acting beyond one’s self interest in accordance with the greater good of humanity. Through using astronomy to highlight our common humanity. 
    \item \textbf{SDG13 (Climate Action):} Take urgent action to prevent climate change and its impacts. Introducing a global perspective of our continually-changing planet as part of a larger solar system, Milky Way Galaxy and an enormous Universe, promulgates the message of the vulnerability of the earth and the human race to climate change.
    \item \textbf{SDG16 (Peace, Justice, and Strong Institutions):} Promote peaceful and inclusive societies for sustainable development, provide access to justice for all and build effective, accountable and inclusive institutions at all levels. Perhaps the most important problem that prohibits achieving peaceful and inclusive societies is conflicts between different ethnic groups within nations and nationalist-based conflicts between countries. In the age of sophisticated weaponry and climate change, such global conflicts not only impede sustainable development, but they also threaten the very survival of the human race. We therefore regard it crucial to promote a culture of peace and non-violence, global citizenship and appreciation of cultural diversity and its contribution to sustainable development through the astronomical perspective.
\end{itemize}
\begin{figure}[H]
    \centering
    \includegraphics[width=0.2\textwidth]{SDG4.png}
    \includegraphics[width=0.2\textwidth]{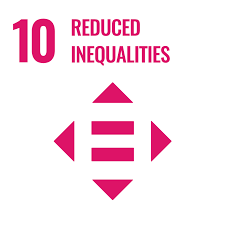}
    \includegraphics[width=0.2\textwidth]{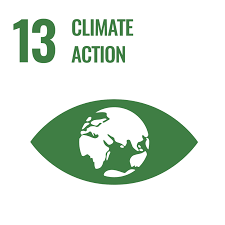}
    \includegraphics[width=0.2\textwidth]{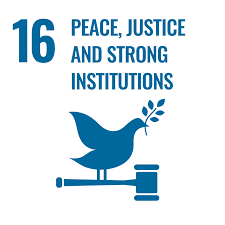}
    \caption{Sustainable Development Goals (SDGs) aligned with astrotourism. Image Credit: \href{https://sdgs.un.org/goals}{United Nations Sustainable Development Goals}\cite{sdgs}}
\end{figure}

\subsection {Theme 3: Knowledge and Skills Development(e.g. data science, data analysis, skills development, hackathons, etc)}
Astro4Skills is focused on both the development of skills through astronomy as well as their application towards sustainable development. It includes skills used in astronomy, such as programming, data handling, data analysis and machine learning, as well as infrastructure such as cloud computing platforms, to advance development objectives through either the teaching or application of skills. This may be executed in the form of programs such as advanced schools/workshops, or in the form of hackathons/competitions which bring together skilled professionals or students with the goal of solving development issues using such skills.

How can astronomy contribute?
The implementation of training schools/workshops could lead to  an increase in the number of students/individuals who are confident in applying the skills learned to their own studies, research or careers, as well as  motivated to further the skills learned, leading to enhanced employability and greater contribution. It could also lead to a growing number of cross disciplinary and cross sector data intensive projects, further partnerships with development and data science organizations and wider use of infrastructure such as high performance and cloud computing, especially in contexts where such tools are used infrequently and could be beneficial.

Projects under this theme align with a number of SDGs, including:
\begin{itemize}
    \item \textbf{SDG 4 (Quality Education):} The theme promotes education through workshops, hackathons, and training in data science, programming, and machine learning – especially in underserved regions.
    \item \textbf{SDG 9 (Industry, Innovation and Infrastructure):} Leveraging astronomy-related infrastructure like cloud computing and high-performance computing supports technological advancement and innovation.
    \item \textbf{SDG 10 (Reduced Inequalities):} Targeting marginalized communities and countries with limited resources helps bridge the digital divide and promote inclusive development.
    \item \textbf{SDG 17 (Partnerships for the Goals):} Collaborations with organizations like DataKind and WHO foster cross-sector partnerships to tackle development challenges using astronomy-based data science.
\end{itemize}

\begin{figure}[H]
    \centering
    \includegraphics[width=0.2\textwidth]{SDG4.png}
    \includegraphics[width=0.2\textwidth]{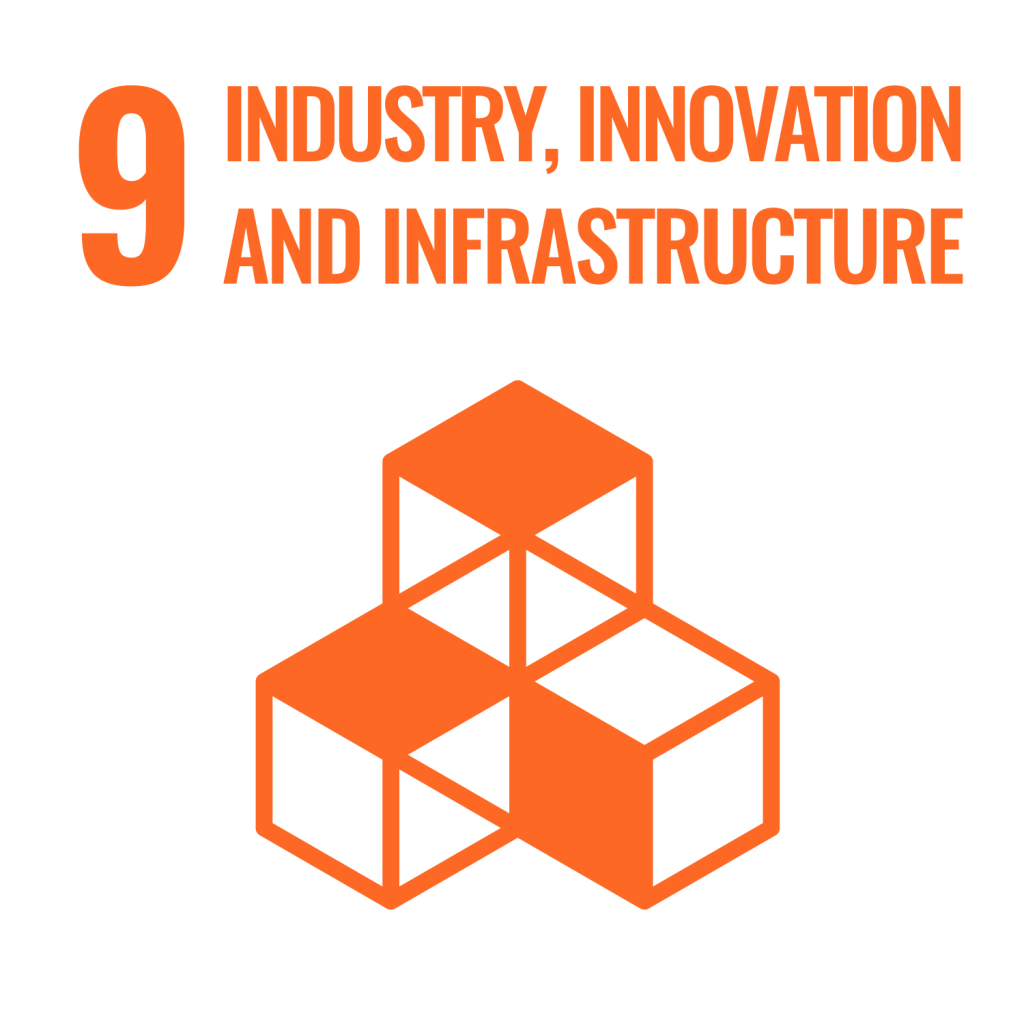}
    \includegraphics[width=0.2\textwidth]{SDG10.png}
    \includegraphics[width=0.2\textwidth]{SDG17.png}
    \caption{Sustainable Development Goals (SDGs) aligned with astrotourism. Image Credit: \href{https://sdgs.un.org/goals}{United Nations Sustainable Development Goals}\cite{sdgs}}
\end{figure}

\subsection{Theme 4: Technology from Astronomy (e.g. software, water, solar, dark skies, etc)}
Astronomy is a powerful driver of technological innovation, generating tools, systems, and methodologies that extend far beyond the study of the Universe. The demands of observing faint and distant objects have led to advances in imaging, data processing, instrumentation, and high-performance computing, many of which have direct applications in addressing real-world development challenges. This theme explores how astronomy-derived technologies and infrastructure contribute to sustainable development, particularly in relation to SDGs 9 (Industry, Innovation and Infrastructure), 7 (Affordable and Clean Energy), 6 (Clean Water and Sanitation), and 13 (Climate Action).

\begin{itemize}
\item \textbf{SDG6 (Clean Water and Sanitation):} Ensure availability and sustainable management of water and sanitation for all. Data processing techniques, remote sensing applications, and imaging technologies originally developed for astronomy are increasingly applied in environmental monitoring, including water resource management, pollution tracking, and climate-sensitive water systems.
\item \textbf{SDG7 (Affordable and Clean Energy):} Ensure access to affordable, reliable, sustainable and modern energy for all. Technologies developed for astronomical research, including advanced solar observation systems and energy-efficient instrumentation, contribute to improvements in solar energy research and optimization. Astronomy-driven innovation also supports remote and off-grid energy solutions used in observatories and rural communities.
\item \textbf{SDG9 (Industry, Innovation and Infrastructure):} Build resilient infrastructure, promote inclusive and sustainable industrialization, and foster innovation. Astronomy has historically driven innovation in areas such as imaging sensors, adaptive optics, data analytics, software development, and high-performance computing. These technologies have wide-ranging applications across health, engineering, environmental monitoring, and digital industries, supporting broader technological capacity building.
\item \textbf{SDG13 (Climate Action):} Take urgent action to combat climate change and its impacts. Astronomy contributes to climate awareness through Earth observation technologies, atmospheric modelling tools, and the promotion of dark sky preservation, which reduces energy waste and light pollution. These contributions support both scientific understanding and practical responses to environmental change.
\end{itemize}

\begin{figure}[H]
    \centering
    \includegraphics[width=0.2\textwidth]{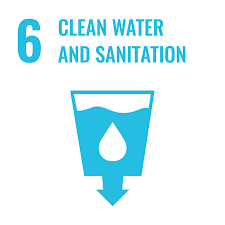}
    \includegraphics[width=0.2\textwidth]{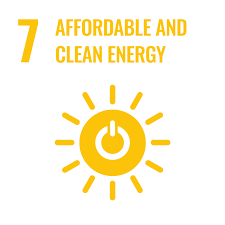}
    \includegraphics[width=0.2\textwidth]{SDG9.png}
    \includegraphics[width=0.2\textwidth]{SDG13.png}
    \caption{Sustainable Development Goals (SDGs) aligned with astrotourism. Image Credit: \href{https://sdgs.un.org/goals}{United Nations Sustainable Development Goals}\cite{sdgs}}
\end{figure}

Overall, astronomy serves as a catalyst for cross-sector technological innovation, with benefits that extend well beyond the discipline itself. By transferring knowledge, tools, and methods from astronomy into applied domains, this theme highlights its role in enabling sustainable, inclusive, and impactful technological development.

\subsection{Theme 5: Addressing Inequality through Astronomy (e.g. gender, geographic, ability, etc)}
Astronomy is a science that belongs to everyone. Regardless of nationality, gender, physical ability, socio-economic status, or geographic location, every person shares the same sky. This universal nature of astronomy provides a unique platform for addressing inequalities and creating opportunities for individuals and communities that have historically been excluded from participation in science, education, and development. The goal of this theme is to leverage astronomy as a tool for inclusion, empowerment, and equitable access to knowledge, skills, and opportunities, contributing primarily to SDGs 4 (Quality Education), 5 (Gender Equality), and 10 (Reduced Inequalities).

\begin{itemize}
\item \textbf{SDG4 (Quality Education):} Ensure inclusive and equitable quality education and promote lifelong learning opportunities for all. Astronomy can inspire curiosity and engagement with science while providing accessible pathways into STEM education. Projects under this flagship seek to expand access to quality educational resources, strengthen science literacy, and increase participation among underserved and marginalized populations.
\item \textbf{SDG5 (Gender Equality):} Achieve gender equality and empower all women and girls. Astronomy can serve as a powerful platform for challenging stereotypes, increasing representation, and creating opportunities for women and girls in science and technology. This flagship promotes initiatives that foster inclusive participation, leadership, and capacity building across all genders.
\item \textbf{SDG10 (Reduced Inequalities):} Reduce inequality within and among countries. Significant disparities persist in access to scientific infrastructure, educational opportunities, and participation in the global scientific community. Through astronomy-based interventions, this flagship aims to reduce barriers related to geography, disability, socio-economic status, and other forms of exclusion, ensuring that the benefits of astronomy are shared more equitably.
\end{itemize}

\begin{figure}[H]
    \centering
    \includegraphics[width=0.2\textwidth]{SDG4.png}
    \includegraphics[width=0.2\textwidth]{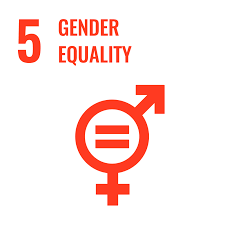}
    \includegraphics[width=0.2\textwidth]{SDG10.png}
    \caption{Sustainable Development Goals (SDGs) aligned with astrotourism. Image Credit: \href{https://sdgs.un.org/goals}{United Nations Sustainable Development Goals}\cite{sdgs}}
\end{figure}

Among these themes, three flagship projects have been identified for global implementation through external fundraising. The Flagship projects fall under the themes: Sustainable, local socio-economic development through Astronomy, Science diplomacy through Astronomy: Celebrating our Common Humanity, and Knowledge and Skills for Development. This approach allows for the scaling of impact of astronomy for development across a significant portion of the world. Each flagship project is discussed in the next section.

\section{Flagship Projects}
\subsection{Astrotourism}
Astrotourism is the intersection of astronomy (the study of objects in outer space) and tourism (visiting places for recreational activities). The aim is to harness the potent synergy of astronomy and tourism to catalyze socioeconomic development at the rural level. Recognizing the inherent advantages of naturally dark skies and rich natural and cultural heritage in many rural areas, the flagship initiative empowers communities (rural, semi-urban and urban), ensuring that the transformative potential of astrotourism becomes an inclusive force, fostering economic growth and preserving local identities.

The Astrotourism Flagship Project, which falls under the socioeconomic development theme (\cite{socioeconomic}), is structured around four key programmatic pillars: 
\begin{itemize}
    \item economic development through astronomy-inspired tourism; 
    \item cultural preservation and social inclusion;
    \item environmental sustainability and dark sky protection;
    \item bridging inequality divides through shared access to the night sky.
\end{itemize}

The Flagship project supports community-driven astro-experiences that create local economic opportunities, celebrate Indigenous and cultural sky knowledge, and promote protection of dark skies and natural environments. At the same time, astrotourism can be seen as a way to connect people across social and economic divides through a shared sense of humanity under the stars.

\subsection{Astronomy for Mental Health}
The Astronomy for Mental Health flagship is focused on harnessing the inspirational potential of astronomy, and using it as a tool for improving people’s mental health and well-being. The overarching goal is to use astronomy to empower individuals and communities to reach their full potential. By utilizing various methods within astronomy, such as stargazing and educational initiatives, the project seeks to create settings for self-exploration and reflection, fostering a profound sense of awe and rejuvenation. Through interdisciplinary collaborations and a focus on accessibility, the initiative promotes cost-effective and culturally relevant astronomy activities. These activities are designed to be easily understandable and memorable, providing an inclusive approach to mental health improvement. The flagship project is aligned with SDG 3, our goal is to reduce premature mortality from non-communicable diseases and strengthen the prevention and treatment of substance abuse. By utilizing astronomy as a tool, we promote mental health and well-being, contributing to the reduction of anxiety, depression, and related disorders. 

The flagship activities offer therapeutic experiences that support overall health and resilience. Dedicated to research, evaluation, and iterative improvement, the flagship engages with mental health professionals and conducts thorough reviews of psychological literature. This ensures that the tools and methods used are grounded in robust psychological practices. 

\subsection{Astronomy for Skills}
The flagship projects under the Astronomy for Skills theme focuses on leveraging astronomy-related expertise to address global development challenges. One key initiative was the Big Data Hackathons, organized by the Office of Astronomy for Development (OAD) and Development in Africa with Radio Astronomy (DARA) Big Data, which aimed to build data science and programming capacity across Africa using radio astronomy as a foundation. These multi-year events have empowered participants with critical technical skills while fostering regional innovation (\cite{bigdata}).

Another major project is Hackathons for Development (Hack4Dev), a joint initiative with the participation of several partners, including the Inter-University Institute for Data Intensive Astronomy (IDIA), the Office of Astronomy for Development (OAD), the BRICS Intelligent Telescope and Data Network (BITDN), DARA Big Data, and the African Astronomical Society (AfAS). This initiative aims to implement a series of hackathons designed to apply astronomy-derived skills – such as machine learning, cloud computing, and data analysis – to real-world development problems (\cite{framework}). Held in countries like Kenya, Ghana, South Africa, India, and Brazil, these events bring together students, professionals, and industry partners to co-create solutions that are both locally relevant and globally impactful.

Together, these projects exemplify how astronomy can be a powerful catalyst for education, economic growth, and social progress. 

\section{The OAD Flagship Ecosystem}
\begin{figure}[H]
    \centering
    \includegraphics[width=0.5\textwidth]{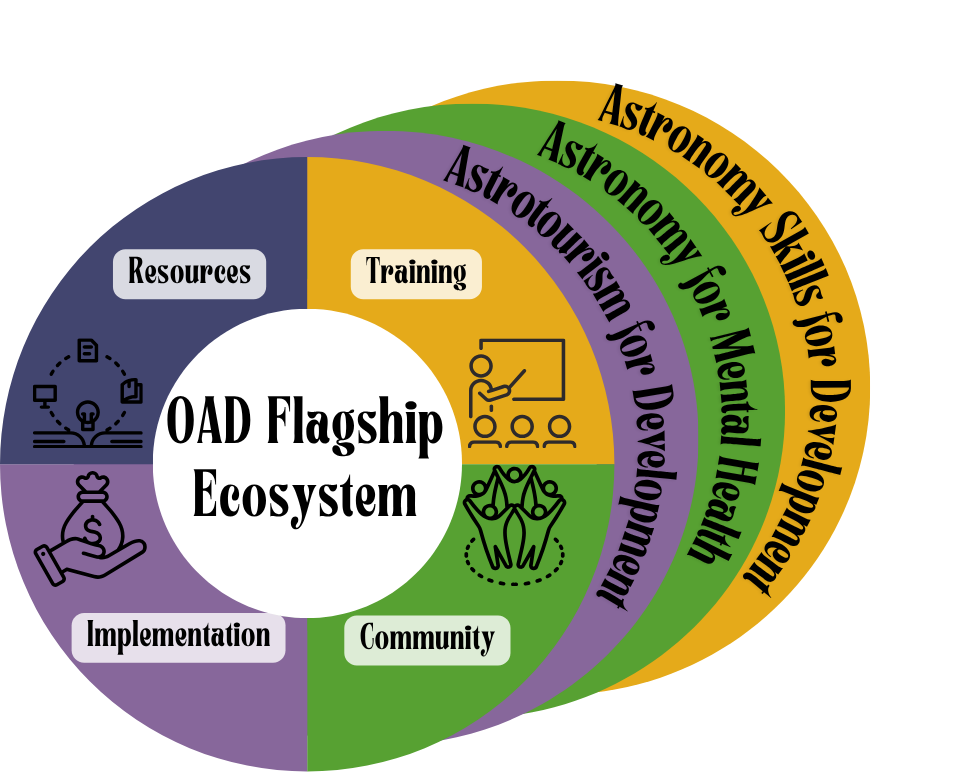}
    \caption{The OAD Flagship Ecosystem.}\label{ecosystem}
\end{figure}

The OAD’s flagship work is organized around four key areas: Resources, Training, Implementation, and Community (see Figure \ref{ecosystem}). Together, these areas form part of a dynamic ecosystem that provides the tools, knowledge, and networks necessary to transform astronomy into a driver of positive change for individuals and communities across the globe.

\subsection{Resources}
The OAD has developed a wide range of openly accessible resources to support individuals, organizations, and institutions in designing and implementing these flagship initiatives across the globe. These resources are intended not only to guide new entrants into the field but also to help existing practitioners broaden their offerings by incorporating astronomy into their activities. 

In addition, the OAD has supported numerous initiatives through its Annual Call for Proposals. Many of these funded projects have generated practical outputs - such as training materials, toolkits, and multimedia content - that are made available for the wider community. Collectively, these outputs contribute to a growing global library of knowledge and practices that can inspire and inform future initiatives.

Case studies emerging from these projects are particularly valuable, as they document lessons learned, highlight innovative approaches, and provide guidance for developing best practices (\cite{socioeconomic}, \cite{casestudies}). In 2020, the OAD published a special issue of the Communicating Astronomy with the Public (\textit{CAP}) Journal, that explored how astronomy can be used as a tool for development, showcasing outreach projects and strategies that promote education, inclusion, gender equality, environmental sustainability, economic growth through astrotourism, and social cohesion (\cite{CAP}). They also demonstrate how astronomy can act as a catalyst for sustainable development. 

Another important aspect of these resources is building the evidence base that supports astronomy-based interventions and their impact on society. Through research, evaluation, and knowledge-sharing, we can better understand how astronomy contributes to sustainable development and how its impact can be strengthened for communities worldwide. This area is still under development, with ongoing research, as well as several publications and preprints already available. Some publications from the OAD include: \cite{comment1}-\cite{mentalhealth} and from project leaders and partners \cite{resource4}-\cite{comment4}.

There are also resources available for the flagship projects designed by the OAD and/or developed in partnership with collaborators. For the Astrotourism flagship project, resources have been developed to facilitate meaningful connections between astronomy and community-led experiences. These resources are designed to leverage the strengths of various stakeholder groups rather than focusing solely on gaps (\cite{resources}). For the Astronomy for Mental Health flagship project, resources have been created to bridge astronomy and mental well-being initiatives (\cite{mentalhealth1}, \cite{narratives}). For the Astronomy for Skills flagship project, resources are available to guide users in hosting their own hackathons, as well as to provide challenges that can be implemented during these hackathons, hosted on \href{https://github.com/Hack4Dev}{GitHub}.

By making these resources openly available, the OAD ensures that communities, practitioners, and decision-makers worldwide can access the tools they need to build initiatives that are impactful, inclusive, and sustainable.

\subsection{Training}
Capacity building is a central component of the flagship projects. The OAD recognizes that in order to foster sustainable growth in this field, it is essential to equip individuals, communities, and institutions with the necessary knowledge and skills.

One of the primary tools for training is free online courses, which introduce participants to the fundamentals of each flagship and provide practical guidance for designing and implementing related initiatives. These courses are hosted on the OpenLearn platform, a free educational platform provided by The Open University. It offers thousands of free courses, articles, videos, and interactive activities across various subjects, allowing anyone in the world to learn at their own pace without paying tuition fees. Upon completion of the courses, users earn free digital badges or Statements of Participation to showcase their achievements. There are currently two flagship courses available (\href{https://astrotourism.astro4dev.org/online-course/}{Astrotourism}\footnote{\href{https://astrotourism.astro4dev.org/online-course/}{Astrotourism} online course} and \href{https://mentalhealth.astro4dev.org/online-course/}{Astronomy for Mental Health}\footnote{\href{https://mentalhealth.astro4dev.org/online-course/}{Astronomy for Mental Health} online course}), as well as an introductory course in \href{https://www.open.edu/openlearncreate/course/view.php?id=3071}{Astronomy for Development}\footnote{\href{https://www.open.edu/openlearncreate/course/view.php?id=3071}{Astronomy for Development} online course} aimed at assisting those interested in submitting a proposal for the IAU OAD annual call for proposals.

Beyond the courses, the OAD also facilitates in-person training delivered in collaboration with its network of partners and regional offices. These sessions combine theoretical instruction with experiential learning, enabling participants to apply their knowledge directly within their own contexts. Training programs can also be co-developed and tailored to specific needs, ensuring relevance to local conditions, cultures, and aspirations. In addition, training efforts are complemented by mentorship and peer-learning opportunities within the Community of Practice, where participants can continue to develop their expertise, share experiences, and access ongoing support.

\subsection{Community}
The Community of Practice is a dedicated space for collaboration, peer learning, and professional exchange. The online platform, hosted on \href{https://discord.gg/QTeSbEteCM}{Discord}\footnote{{Anyone is welcome to join our \href{https://discord.gg/QTeSbEteCM}{Discord} server}}, brings together diverse stakeholders, including astronomers, tourism operators, rural community members, educators, cultural practitioners, mental health professionals, and enthusiasts. By convening this broad network, the community fosters dialogue across disciplines and regions, creating opportunities for innovation and shared growth.

Members exchange resources, best practices, and case studies, while also collaborating on joint initiatives that promote sustainability and inclusivity. The platform serves not only as a hub for knowledge-sharing but also as a support system where members can seek guidance, mentorship, and inspiration from peers with similar interests and challenges.

Importantly, the community emphasizes ethical and sustainable practices, encouraging members to follow evidence based practices, integrate local knowledge, respect cultural heritage, and minimize environmental impacts in their work. By nurturing these values, the Community of Practice helps ensure that astronomy acts as a tool for empowerment and development.

As part of its community activities, the OAD hosts online events focused on its flagship themes. These engagements aim to encourage community participation and the exchange of knowledge and resources. They play an important role in fostering meaningful discussions and generating valuable insights within the community (\cite{cop}).

\subsection{Implementation}
Implementation is primarily achieved through the IAU OAD Annual Call for Proposals, which serves as a central mechanism for identifying, supporting, and scaling innovative projects around the world. Through this process, individuals and organizations are guided in the development of their proposals, benefiting from resources and advice provided by the OAD and its partners. Once selected, projects receive not only financial support in the form of seed funding, but also structured assistance in monitoring and evaluation to ensure accountability, effectiveness, and long-term impact. This systematic approach helps transform promising ideas into sustainable initiatives capable of creating lasting benefits for local communities (\cite{monitoring}, \cite{impactcycle}).

Beyond the Annual Call, implementation is further supported by the OAD’s Regional Offices, which act as key facilitators of local engagement. These offices provide regional expertise, contextual understanding, and ongoing support to projects, ensuring that activities are rooted in the cultural, social, and economic realities of the communities they serve. 

These flagship projects represent some of the most effective and successfully tested applications of astronomy for development. The OAD has also established Flagship Funds to raise support for the global implementation of these initiatives.

\section{Conclusion}
The Flagship Ecosystem was designed not only to support the current OAD flagship projects but also to provide a flexible and scalable framework for future astronomy-for-development initiatives. Built around four interconnected pillars: Resources, Training, Community, and Implementation, the ecosystem seeks to lower barriers to participation while enabling the sustainable growth of astronomy-based interventions around the world. It is grounded in key principles of science for development, including inclusivity, openness, sustainability, societal impact, evidence-informed practice, innovation and the knowledge economy, and participatory approaches. By integrating these principles into a coherent support structure, the ecosystem strengthens the ability of individuals, organizations, and communities to translate ideas into meaningful action.

More broadly, the Flagship Ecosystem represents a shift from funding individual projects towards cultivating a global ecosystem for Astronomy for Development. One that supports learning, collaboration, evidence generation, and long-term impact. By combining open resources, capacity building, communities of practice, and implementation support, it creates pathways for successful initiatives to be replicated, adapted, and scaled across diverse contexts. Together, these structures enable the translation of global strategies into practical, locally relevant outcomes, ensuring that astronomy can continue to contribute meaningfully to sustainable development worldwide.

\section{Acknowledgements}
The Office of Astronomy for Development is a joint project of the International Astronomical Union (IAU) and the South African National Research Foundation (NRF) with the support of the Department of Science, Technology, and Innovation (DSTI).

\end{document}